\documentstyle[12pt]{article}
\def\beq{\begin{equation}}
\def\eeq{\end{equation}}
\def\etapigg{\eta  \rightarrow \pi^{0} \gamma \gamma}
\def\ggpipi{\gamma \gamma \rightarrow \pi^{0} \pi^{0}}

\begin{document}

\begin{titlepage}
\begin{flushright}
UMN--TH--1219/93  \\ TPI--MINN--93/50--T \\  SNUTP--93--72
\end{flushright}
\vspace{0.2in}
\begin{center}
{\Large \bf   $\etapigg$ and $\ggpipi$ in  \\
$O(p^{6})$ chiral perturbation theory \\}
\vspace{0.6in}
{\bf Pyungwon Ko
\footnote{e-mail : pko@phyb.snu.ac.kr }
} \\
\vspace{0.4in}
{\sl
Department of Physics  \\
Hong-Ik University  \\
Seoul 121-791, KOREA
\\}
\vspace{0.4in}
{\bf   Abstract  \\ }
\end{center}
$\eta \rightarrow \pi^{0} \gamma  \gamma$ and
$\ggpipi$  are considered in $O(p^{6})$ chiral perturbation theory.
In addition to the usual $\rho,\omega$ contributions,  there are two
$O(p^{6})$  operators  (${\cal L}_{6,m}$) arising  from explicit chiral
symmetry breaking.   Since only one of the two operators contributes to
$\etapigg$, the coefficient of this  operator ($\equiv d_3$)   can be
determined in two ways : (i) from  the measured decay rate of  $\eta
\rightarrow \pi^{0} \gamma \gamma$, and (ii) by assuming the resonance
saturations of the low energy coefficients in the $O(p^{6})$ chiral
lagrangian. We find that two methods lead to vastly different values of
$d_3$, which would indicate that either the measured decay rate for
$\eta \rightarrow \pi^{0} \gamma \gamma$ is too large by a factor of
$2 \sim 3$, or the resonance saturation assumptions do not work for
$d_3$.

\vspace{.3in}
\noindent
%PACS numbers : 13.10.+q,\ 13.35.+s,\
\end{titlepage}

\baselineskip=20pt

1.  The decay  $\eta \rightarrow \pi^{0} \gamma \gamma$ is interesting
both theoretically and experimentally.
It is  a potential background to the decay $\eta \rightarrow
\pi^{0} e^+ e^-$, which can be a sensitive probe for $C-$violation
in hadronic electromagnetic interactions and is an important topic
at the Saturne eta facility \cite{nefkens}.
To search for any $C-$violating effect in $\eta \rightarrow \pi^{0}
e^+ e^-$, one has to know  the contribution of the $C-$conserving
two--photon process, $\eta \rightarrow \pi^{0} \gamma^* \gamma^*
\rightarrow \pi^{0} e^+ e^-$.
Thus, one needs better understanding of  $\etapigg$.

If one computes the decay rate of $\etapigg$ in  $O(p^{4})$ chiral
perturbation theory, the result comes out to be too small ($\approx 3.9
\times 10^{-3}$ eV) \cite{ametller} compared to the measured value
\cite{pdg},
\beq
\Gamma_{exp} (\etapigg) = ( 0.84 \pm 0.18 ) ~~{\rm eV}.
\label{eq:gexp}
\eeq
Therefore, we are forced  to consider higher order ($ \geq O(p^{6})$)
contributions.   The usual $\rho$ and $\omega$ exchanges \cite{cheng}
can explain only a part of the observed decay rate \cite{ametller},
and it remains unclear which gives additional contributions in chiral
perturbation theory.   Whatever the answer may be, we have to keep in mind
that any explanations for $\etapigg$ should not lead us to wrong
descriptions of $SU(3)-$related processes such as the electric and
magnetic polarizabilities of $\pi^0$ and  $\ggpipi$.

In this paper, we study  $\etapigg$ and $\ggpipi$ in $O(p^{6})$ chiral
perturbation theory.  We first summarize the work of Ametller {\it et al.}
\cite{ametller}, where two $O(p^{6})$ operators were considered at tree level
and assumed to be saturated via $\rho, \omega$ and other resonances.
Then, we point out that there are two more
operators which are effectively relevant to these processes at the
same order in the chiral expansion.  These two operators (called $d_{3}$ and
$d_4$ terms below) are proportional
to explicit chiral symmetry breaking  due to nonvanishing  current quark
masses.   Since only one of these two (say, the $d_3$ term) contributes to the
decay $\etapigg$,  one can  determine the coefficient $d_3$
from the measured decay rate of $\etapigg$.
There are two solutions for $d_3$, for which
we  predict the $m_{\gamma \gamma}^2$ spectrum in $\etapigg$.
For the resulting values of $d_3 $,  the $d_3$  term  contributes too much
to the electromagnetic polarizabilties of $\pi^0$
(or, equivalently,  gives wrong predictions on the reaction
$\ggpipi$).  However, this can be cancelled  by a suitable choice of
$d_4$.  Thus, having two independent operators, one can fit both $\etapigg$
and $\ggpipi$ without any problems.
Then, we consider resonance (more specifically, $a_{0} (980)$)
contributions to $d_3$, and compare with numerical values of $d_3$'s
determined by (1).  We find these two are vastly different from each other,
and discuss its implications.
The results are summarized at the end.

Before delving into main discussions, let us define the amplitude for
$\etapigg$  in the following way :
\begin{eqnarray}
{\cal M} ( \eta(p) \rightarrow \pi^{0} (p^{'}) \ \gamma (k,\epsilon) \
\gamma (k^{'},\epsilon^{'}) ) = \epsilon^{\mu} \epsilon^{' \nu} ~
\left[  a(s,t) \left( g_{ \mu\nu} k \cdot k^{'} - k_{\mu}^{'} k_{\nu}
\right)   \right.   \nonumber
\\
\left. + b(s,t) \left( - g_{\mu\nu} p \cdot k p \cdot k^{'} - k\cdot k^{'}
p_{\mu} p_{\nu} + p\cdot k k_{\mu}^{'} p_{\nu} + p\cdot k^{'} k_{\nu}
p_{\mu}   \right) ~\right],
\label{eq:form}
\end{eqnarray}
where
\[
s = (p - p^{'})^{2} = ( k + k^{'} )^{2} \equiv m_{\gamma\gamma}^{2},
\]
\[
t = (p - k )^{2}, ~~~~~~~u = ( p - k^{'} )^{2}.
\]
The amplitude for $\ggpipi$ can be obtained from (\ref{eq:form}) by
crossing symmetry with $p^{2} = m_{\pi}^2$.

The form factors $a(s,t)$ and $b(s,t)$ will be calculated in the framework
of chiral perturbation theory \footnote{For the calculations based on the
quark model, see \cite{ng}.
%and \cite{dawson}.
}.
For the processes under considerations, the usual $3\times 3$
matrix field $U(x)$ for the Nambu--Goldstone bosons is simplified as
$ U_{0} (x) \equiv e^{2 i M_{0} (x) /f_{\pi}}$,
where $f_{\pi} = 93$ MeV is the pion decay constant and
\begin{equation}
M_{0} \equiv {1\over \sqrt{2}}~{\rm diag} ~( {\pi^{0} \over \sqrt{2}}
+ {\eta \over \sqrt{3}}, - {\pi^{0} \over \sqrt{2}} +
{\eta \over \sqrt{3}},  {-  \eta \over  \sqrt{3}} ).
\end{equation}
We have used the nonet symmetry to include a $SU(3)$ singlet $\eta_1$,
and the physical $\eta$ is defined as
\begin{equation}
\eta = \eta_{8} \cos\theta_{p} - \eta_{1} \sin\theta_{p}
\approx {1\over \sqrt{3}}~( u \bar{u} + d \bar{d} - s \bar{s} ),
\end{equation}
with $\theta_{p} \simeq - 20^{\circ}$.
Therefore, the $U_{0} (x)$ commutes with both the quark mass matrix
$\mu m = \mu (m_{u}, m_{d}, m_{s})$
and the quark electric charge matrix $Q = {\rm diag} (2/3,-1/3,-1/3)$.
This property is useful when we construct $O(p^{6})$ operators in the
following.

\vspace{.3in}

2.  Let us first consider $\etapigg$.  Study of this decay in
chiral perturbation theory is described in detail in Ref.~\cite{ametller}.
The tree level amplitudes at $O(p^{2})$ and $O(p^{4})$ are zero, since
$[ Q, U_{0} ] = 0$.   The $O(p^{4})$ loop amplitudes
due to pions and kaons are obtained explicilty, and their contributions are
very small because of the $G-$parity and the
largeness of kaon mass compared to the available $m_{\gamma\gamma}$.
One can expect this continues to be true for the $O(p^{6})$ loop amplitude
for $\etapigg$ \cite{ametller}.
Therefore, the decay $\etapigg$ is rather a unique
process in chiral perturbation theory in the sense that it receives
main contributions from the $O(p^{6})$ tree level chiral lagrangian.

There would be many operators in  the tree level chiral lagrangian at
order  $O(p^{6})$,  which serve as counterterms to absorb divergences
appearing in the loop amplitudes at the same order in the chiral
lagrangian.  Therefore, it would be impossible to write down all
dimension--6 operators and determine their coefficients
from phenomenology of light pseudoscalar mesons.
However, the situation becomes considerably simplified, if we consider only
normal two--photon processes of neutral mesons such as  $\etapigg$ and
$\gamma \gamma \rightarrow \pi^{0} \pi^{0}$.  The reason is that we can
consider only $U_{0}$ which enjoys the property
\beq
[ U_{0}, Q ] = [ U_{0}, \mu m ] =0.
\eeq
At $O(p^{6})$, we construct operators with (i) two $Q$'s and (ii) two
derivatives on $U_{0}$ or one insertion of $\mu m$, in order to describe
$\etapigg$ and $\ggpipi$.

The authors of Ref.~\cite{ametller} considered   two $O(p^{6})$ operators,
keeping only terms with single traces over flavor indices by invoking the
large $N_c$ limit :
\begin{equation}
{\cal L}_{6}  =  d_{1}~F_{\mu\nu}F^{\mu\nu}~{\rm Tr} \left[  Q^{2}
\partial_{\alpha} U_{0} \partial^{\alpha} U_{0}^{\dagger} \right]
+ d_{2}~F_{\mu\alpha}F^{\mu\beta}~{\rm Tr} \left[ Q^{2} \partial^{\alpha}
U_{0}^{\dagger} \partial_{\beta} U_{0} \right].
\label{eq:l6p}
\end{equation}
Then, the coefficients $d_{1,2}$ were determined assuming that these are
saturated by suitable resonance exchanges.

The most important term corresponds to  $\rho$ and $\omega$ exchanges
\cite{cheng} :
\begin{eqnarray}
a_{V} (s,t) & = & {2\sqrt{2} \over 3\sqrt{3}}~{g_{\omega\pi^{0}\gamma}^{2}
\over 2}~\left[ { t + m_{\eta}^{2} \over t - m_{\rho}^2} +
 { u + m_{\eta}^{2} \over u - m_{\rho}^2} \right],
\label{eq:av6eta}
\\
b_{V} (s,t) & = & {2\sqrt{2} \over 3\sqrt{3}} ~g_{\omega\pi^{0}\gamma}^{2}
{}~\left[ {1 \over  t - m_{\rho}^2} + { 1 \over  u - m_{\rho}^2} \right],
\label{eq:bv6eta}
\end{eqnarray}
where $g_{\omega\pi^{0}\gamma} = 0.71~{\rm GeV}^{-1}$ from $\omega \rightarrow
\pi^{0} \gamma$.   If the full propagator is used for the intermediate
vector mesons,   one gets  $\Gamma_{\rho + \omega} (\etapigg) = 0.31 ~~
{\rm  eV}$.  Combining with the $O(p^{4})$ loop amplitude, one gets
$0.36$ eV, which is less than a half of the measured width (\ref{eq:gexp}).
Including the $O(p^{8})$ amplitude with  two vertices  from the Wess--Zumino
anomaly,  one gets a little improvement, $0.42$ eV.
One can add more resonances such as $a_{0}, a_{2}$ \cite{ametller} or
$b(1235), h(1170), h(1380)$ \cite{ko1}, but these contributions are still
small and/or have sign ambiguities relative to the $\rho,\omega$
contributions.  This is the current status of  $\etapigg$ in chiral
perturbation theory.

Following the spirit of chiral perturbation theory, however, other $O(p^{6})$
operators should be considered as well as (\ref{eq:l6p}).
First of all, there can be terms with two derivatives and double traces
which were ignored in Ref.~\cite{ametller}.  These terms contribute to
$\etapigg$ and $\ggpipi$ with different weights from those of (\ref{eq:l6p}).
When one calulate the $\rho,\omega$ contributions to $\ggpipi$,
the result is consistent with assuming the validity of (\ref{eq:l6p}).
Therefore, we still ignore such terms with two derivatives and double traces.
The remaining possibilities with one insertion of $\mu m$ are
\begin{eqnarray}
{\cal L}_{6,m} & = & d_{3} F_{\mu\nu} F^{\mu\nu}~{\rm Tr}
\left[ \left\{ Q^{2}, \mu m
\right\} ~( U_{0} + U_{0}^{\dagger} ) \right]   \nonumber  \\
& + & d_{4} F_{\mu\nu} F^{\mu\nu}~{\rm Tr} [ Q^{2} ]~
{\rm Tr} \left[  \mu m ( U_{0} + U_{0}^{\dagger} ) \right].
\label{eq:l6m}
\end{eqnarray}

It is straightforward to calculate the contributions of
$O(p^{6})$ chiral lagrangians, (\ref{eq:l6p}) and (\ref{eq:l6m}),
to  $\etapigg$ :
\begin{eqnarray}
a_{6} (s,t) & = & {2 \sqrt{2}  \over 3 \sqrt{3} f_{\pi}^{2}}~\left[
- 4 d_{1} p \cdot p^{'} + d_{2} p \cdot ( k + k^{'} ) + 4 d_{3} m_{\pi}^{2}
\right], \\
b_{6} (s,t) & = & {4 \sqrt{2} \over 3 \sqrt{3} f_{\pi}^2}~d_{2}.
\end{eqnarray}
Since we have assumed that $\omega$ and $\rho$ exchanges dominate $d_1$ and
$d_2$, there remains only one unknown parameter, $d_3$, which can be
fixed from the observed decay rate of  $\etapigg$, (1).
Including the $O(p^{4})$ loop amplitude and the $\rho,\omega$ exchanges,
one finds that
\beq
d_{3} = ( - 1.4 \pm 0.4 ) \times 10^{-2} ~~{\rm GeV}^{-2}~~~~{\rm or}~~~~
( 4.5 \pm 0.4 ) \times 10^{-2}~~{\rm GeV}^{-2}.
\label{eq:d3}
\eeq
(Even if we include the $O(p^{8})$ amplitude obtained in Ref.~\cite{ametller},
the results do not change very much :
\[
d_{3} = ( - 1.2 \pm 0.4 ) \times 10^{-2} ~~{\rm GeV}^{-2}~~~~{\rm or}~~~~
( 4.7 \pm 0.4 ) \times 10^{-2}~~{\rm GeV}^{-2},
\]
which are consistent with (\ref{eq:d3}) within $1 \ \sigma$ level. )
Compared with the vector meson contributions to $d_{1,2}$,
\beq
d_{1}^{\omega+\rho} = - {1\over 2}~d_{2}^{\omega+\rho} = {g_{\omega\pi^{0}
\gamma}^{2} f_{\pi}^{2} \over 2 m_{\rho}^2}~ \simeq 0.37 \times 10^{-2}~~
{\rm GeV}^{-2},
\eeq
the $d_3$'s which will reproduce the measured decay rate (1) are an order of
larger.  This may be annoying in the spirit of chiral perturbation theory.
Presumably, every low energy constant should be of the same order of
magnitude.  This issue will be discussed more, when we consider resonance
contributions to $d_3$.

At the moment, we accept these two values of $d_3$'s for which (1) is
satisfied, and show the $m_{\gamma \gamma}^{2}$ spectrum
in Fig.~1 for each $d_3$.  The solid and the dashed curves correspond to
$d_{3} = -1.4 \times 10^{-2} ~~{\rm GeV}^{-2}$ and $ 4.5 \times 10^{-2}
 ~~{\rm GeV}^{-2}$, respectively.    For comparison, the vector meson
contributions with $d_{3} = 0$ is shown in the dotted curve with the
overall normalization scaled to yield the observed decay rate.
The number of events at low and high $m_{\gamma\gamma}^2$ regions are
different for two values  of $d_3$'s.
Since the quark model calculations for $\etapigg$ \cite{ng}
lead to a different $b(s,t)$ form factor than ours,  the $m_{\gamma\gamma}^2$
spectrum can distinguish the quark model calculation from ours in principle.
Therefore, it is highly desirable to have  good measurements of
the $m_{\gamma\gamma}^2$ spectrum as well as of the absolute decay rate
for $\etapigg$.  This may be achieved at the Saturne eta tagging facilty.

\vspace{.3in}

3.   At this stage, it is important to check that the $d_3$ term
does not contribute too much to the polarizabilities of $\pi^0$, or
equivalently, to the reaction $\ggpipi$.   The electric and the magnetic
polarizabilities can be
obtained from Compton scattering or $\ggpipi$ \cite{donoghue1}.
Since  Compton scattering of $\pi^0$ is not feasible,
one has to resort to the reaction $\ggpipi$ which determines $(\alpha_{E} -
\beta_{M})^{\pi^0}$.  This subject is rather involved because of the final
state $\pi\pi$ interaction.
It is essential to include partial wave unitarity of  elastic $\pi\pi$
scattering  correctly \cite{truong}.
%The pion rescattering model \cite{rosner}
%well describes  this process   up to $\sqrt{s} \sim 900$ MeV.
%%If the phase shifts of the $\pi\pi$ system
%and the  vector meson exchange are taken
%into  account in the dispersion approach, this reaction below $\sqrt{s} <
%700$ MeV  is well understood \cite{truong}.
Using the Crystal Ball data on $\ggpipi$, Kaloshin and Serebryakov
extracted   \cite{kaloshin}
\begin{equation}
( \alpha_{E} - \beta_{M} )^{\pi^0} = (0.8 \pm 2.0) \times 10^{-4} ~~{\rm
fm}^{3}.
\label{eq:kaloshin}
\end{equation}

One can calculate the contributions of (\ref{eq:l6p}) and (\ref{eq:l6m})
to $\ggpipi$.
Again assuming that $d_1$ and $d_2$ terms are dominated by the
$\rho$ and $\omega$ exchanges, we  find
\begin{eqnarray}
a_{6} (s,t) & = & a_{6}^{V} (s,t) + {8 m_{\pi}^{2} \over 9
f_{\pi}^2}~(5 d_{3} + 6 d_{4} )   \\
b_{6} (s,t) & = & b_{6}^{V} (s,t).
\end{eqnarray}
where $a_{6}^{V}$ and $b_{6}^{V}$ can be obtained from (\ref{eq:av6eta}) and
(\ref{eq:bv6eta}) by replacing $2\sqrt{2}/3\sqrt{3} \rightarrow 10/9$ and
$m_{\eta}^{2} \rightarrow m_{\pi}^2$ \cite{ko}.
It is well known that the $O(p^{4})$ amplitude \cite{donoghue} and
the vector meson exchange amplitude contribute to the electric and the
magnetic polarizabilities by \cite{donoghue1}
\begin{eqnarray}
\alpha_{E}^{\pi^0} & = & - 0.5 \times 10^{-4}~~ {\rm fm}^{3},
\\
\beta_{M}^{\pi^0} & = & + 1.3 \times 10^{-4}~~ {\rm fm}^{3}.
\end{eqnarray}
The $d_3$  and $d_4$ terms modify the $a(s,t)$ form factor only. Thus,
they do  not change $( \alpha_{E} +  \beta_{M} )^{\pi^0}$, whose nonzero
value is due to the vector meson  exchanges as discussed in Ref.~
\cite{donoghue1}.  However, they can generate a potentially large change in
$( \alpha_{E} - \beta_{M} )^{\pi^0}$ :
\begin{equation}
\delta ( \alpha_{E} - \beta_{M} )^{\pi^0} = 4.2 \times 10^{-2} ~
( d_{3} + 1.2~d_{4} ) ~~{\rm fm}^{3},
\label{eq:aminusb}
\end{equation}
which is $-5.9 \times 10^{-4}~~{\rm fm}^{3}$ for $d_{3} = -1.4 \times 10^{-2}
{}~~{\rm GeV}^{-2}$ and $d_{4} = 0$.  Without the $d_{4}$ term, the $d_3$ term
(which may explain the observed decay rate of $\etapigg$) would induce too
large a change in (\ref{eq:aminusb}), which seems to be contradictory with
the value (\ref{eq:kaloshin}).  The $d_4$ term  can provide cancellation of
such a dangerous change induced by the $d_3$ term.   Thus, the $d_4$ term
is important in order that we can  explain the decay rate of $\etapigg$
and the polarizabilities
of $\pi^0$  simultaneously in $O(p^{6})$ chiral perturbation theory.
Although we cannot determine $d_{4}$ from (\ref{eq:kaloshin}),  it is
very likely that the effect of explicit chiral symmetry breaking in
$\ggpipi$ is very small,  considering the success of current
approaches (which set the second term equal to zero) in  describing
$\ggpipi$ \cite{truong}.

\vspace{.4in}

4. Up to now, we have assumed that bulk of the decay rate of $\etapigg$
is given by explicit chiral symmetry breaking terms of $O(p^{6})$ (the
$d_{3}$ term in (9)), and found two solutions for $d_{3}$, (13).
In this subsection, we investigate if these two values of $d_3$ can be
understood in terms of the resonance saturations of the low energy constants
in the $O(p^{6})$ chiral lagrangian. More specifically, we consider $a_{0}
(980)$ contribution to $\etapigg$,  the lightest resonance that is relevant
to this decay.

The lowest order chiral lagrangian describing interactions between $a_0$
and pion octet is given by \cite{ecker}
\begin{equation}
{\cal L} (a_{0} \pi^{0} \eta) = {2 \sqrt{2} \over \sqrt{3} f_{\pi}^2}~
\left( c_{d} \vec{a}_{0} \cdot \partial_{\mu} \vec{\pi}   \partial^{\mu}
\eta - c_{m} m_{\pi}^{2} \vec{a}_{0} \cdot \vec{\pi} \eta \right).
\label{eq:a0etapi}
\end{equation}
{}From $\Gamma_{exp} = (57 \pm 11)$ MeV \cite{pdg}, we get a constraint :
\begin{equation}
\left| c_{d} ( E_{\pi} - {m_{\pi}^{2} \over m_{a_0}} ) + c_{m}
{m_{\pi}^{2} \over m_{a_0}} \right| = \left| 0.327 c_{d} + 0.019 c_{m}
\right| = 1.12 \times 10^{-2}~~{\rm GeV}^{2}.
\label{eq:cdcm}
\end{equation}
The authors of  Ref.~\cite{ecker} assumed that the
low energy constants ($L_{i}$'s) in $O(p^{4})$ chiral lagrangian would
be saturated by the low lying resonances such as $\rho, a_{0}, \eta^{'}$,
etc., and  obtained
\begin{eqnarray}
|c_{d}| & = & 3.2  \times 10^{-2}~~{\rm GeV},
\nonumber
\\
|c_{m}| & = & 4.2  \times 10^{-2}~~{\rm GeV},
\label{eq:ecker}
\\
c_{d} c_{m} & > & 0.
\nonumber
\end{eqnarray}
These values satisfy the constraint (\ref{eq:cdcm}), and reproduce the
width for $a_{0} \rightarrow \eta \pi^0$ rather well, $\Gamma ( a_{0}
\rightarrow \eta \pi^0 ) =   59~~{\rm MeV}$.

Defining the $a_{0} \gamma \gamma$ coupling  as
\begin{equation}
{\cal L} (a_{0} \gamma \gamma) = g_{s} F_{\mu\nu} F^{\mu\nu},
\label{eq:a0gg}
\end{equation}
and using $\Gamma ( a_{0} \rightarrow \gamma \gamma ) = ( 0.51 \pm 0.26
)~~{\rm keV}$ and $\sigma ( \gamma \gamma \rightarrow a_{0} \rightarrow
\eta \pi^{0})\approx 30~~{\rm  nb}$, one can determine   $g_s$ :
\[
|g_{s}| = 2.6 \times 10^{-3}~~{\rm GeV}^{-1}.
\]

Now, the  $a_0 (980)$ contribution to $\etapigg$ can be readily obtained
from (\ref{eq:a0etapi}) and (\ref{eq:a0gg}) \footnote{In Ref.~
\cite{ametller}, only the $c_d$ term was considered, and found negligible.} :
\begin{eqnarray}
\delta_{a_0} a(s,t) & = & { 8 \sqrt{2} \over 3 f_{\pi}^2}~{g_{s}
( c_{d} p \cdot p^{'} - c_{m} m_{\pi}^{2} ) \over m_{a_0}^2 },
\\
\delta_{a_0} b(s,t) & = & 0.
\end{eqnarray}
The decay rate for $\etapigg$ is a quadratic function of $c_d$ if we use
(\ref{eq:cdcm}).  Including the contributions by the $O(p^4)$ chiral loop,
the vector mesons ($\rho, \omega$) and the $a_0 (980)$,  we  find that
the observed decay rate for $\etapigg$ (1) is  reproduced, if
\begin{eqnarray}
(c_{d}, c_{m}) & = & (-0.11,2.48), (0.45, -7.16),     \label{eq:cdcm1}
\\
               & {\rm or} & (-0.16, 2.16), (0.40,-7.47).   \nonumber
\end{eqnarray}
These values of $(c_{d}, c_{m})$ are vastly different from (larger by an
order of magnitude than) the results of Ecker {\it et al.},
(\ref{eq:ecker}).  Incidentally,  (\ref{eq:cdcm})
yields  $\Gamma (\etapigg) = 0.34$ eV or  0.37 eV.

A few  remarks are in order.
These values (\ref{eq:cdcm1}) correspond to two solutions for $d_3$
obtained in the previous subsection, once  the $c_d$ (which is small)
is neglected.  Also, it should be noted that the reaction $\ggpipi$ is
not affected at all by the $a_0$ resonance.
Viewing our results  on the $a_0 (980)$ contribution to $\etapigg$, one can
conclude either (i) the assumption made in Ref.~\cite{ecker} is invalid for
$L_{5,8}$'s, or (ii) the present data (1) might be too large
by a  factor of $2 \sim 3$. At this point, it may be helpful to remind that
the current data (1) is based on one experiment, in which the background
estimates are  difficult \footnote{I thank the referee for this comment.}.
It is not clear which would be the  correct interpretation at the moment,
although the first one may be less  likely. (Other low energy  constants
$L_{9,10}$ are well saturated by $\rho$ and $a_1$ \cite{ecker}.)
Better measurement of the decay rate and
the photon spectrum in $\etapigg$  will resolve these two alternatives.
If the decay rate stays high around  the current data (1), one has to
abandon the  assumption that the low
energy constants in $O(p^4)$ chiral lagrangian are saturated by the
low lying resonances for $L_{5,8}$.   Turning around our arguments,
the decay rate for $\etapigg$ would be around $0.42 $eV as estimated by
Ametller {\it et al.} if the $L_{5,8}$'s are   saturated by the scalar
octet containing $a_{0} (980)$.     Thus, it is quite important
to have an improved measurement of $\Gamma (\etapigg)$ at the Saturne
$\eta$ facility.

\vspace{.4in}

5. In conclusion, we considered the decay $\etapigg$  and  polarizabilties
of $\pi^0$ (or, $\ggpipi$)  in $O(p^{6})$ chiral perturbation theory.
For these processes, there are two types of $O(p^{6})$ operators,
(\ref{eq:l6p}) and  (\ref{eq:l6m}).  As usual, (\ref{eq:l6p}) is assumed
to be dominated  by the  $\rho$ and $\omega$ exchanges.
On the other hand, (\ref{eq:l6m}) represents  the effects
of explicit chiral symmetry breaking,  only one of which contributes to
$\etapigg$.
The coefficient of this operator, $d_3$, was determined from the observed
decay rate of $\etapigg$, and the two--photon spectrum was predicted for
two solutions of $d_3$.   Although the $d_3$ term alone does induce
a large amount of $(\alpha_{E} - \beta_{M})^{\pi^0}$,  the other operator
(the $d_4$ term) can provide cancellations to reproduce a value consistent
with (\ref{eq:kaloshin}).
Then, we estimated the low energy constant $d_3$ assuming the resonance
saturation approximation is valid as in $L_{9,10}$ of $O(p^{4})$ chiral
lagrangian.  It turns out that the resulting $d_{3}$ is too small,
yielding 0.34--0.37 eV.
Thus, we end up with a decay rate problem for $\eta \rightarrow \pi^{0}
\gamma \gamma$.
This would imply that either (i) the resonance saturation assumption is
not valid for $L_{5,8}$ and/or for $d_{1,2,3}$ or (ii) the experimental
vaue (1) is too large by a factor of $2 \sim 3$.  Although the resonance
saturation assumption works fine with $O(p^{4})$ chiral lagrangian, there is
no solid evidence that it also works for $O(p^{6})$ chiral lagrangian.
Therefore, we prefer to keep the case (i) as a logical possibility to
the decay rate  problem of $\eta \rightarrow \pi^{0} \gamma \gamma$.
In this sense, a better measurement of decay rate of $\eta \rightarrow
\pi^{0} \gamma \gamma$ is not only imporatant by itself, but also test
the resonance saturation assumption for the low energy constants appearing
in $O(p^{4,6})$ chiral  lagrangians.
Also, the measurement of the $ m_{\gamma \gamma}^2 $ spectrum in $\etapigg$
could test and distinguish various models for this decay, and thus reserve
to further study.
All of these could be achieved  at the Saturne eta  tagging facility.

\vspace{.5in}

{\Large \bf  Acknowledgements}

\vspace{.3in}
Part of this work was done at the  Center for
Theoretical Physics, Seoul National University.    The author
thanks  Prof. H.S. Song and Prof. Jihn E. Kim for their
hospitality during  his stay at CTP.   He is also  grateful to Prof. J.L.
Rosner for  reading the manuscript and making comments, and to the referee
for valuable suggestions (which has been newly incorporated in the
subsection 4).   This work was supported
in part by Department of Energy  grant \# DOE--DE--AC02--83ER--40105, and
by KOSEF through CTP.

\vspace{.5in}

{\Large \it Note Added}

While this work is being revised, a new preprint has appeared \cite{new},
 in which
the same processes were considered in the extended Nambu Jona--Lasinio
model.  In Ref.~\cite{new}, the large $N_c$ limit was taken so that the
$d_4$ term in our lagrangian, (9), is not considered.
Including the constituent quark loops which is absent in the usual chiral
lagrangian approach, they found that $d_{3} \approx 2 d_{3,res}$ and
$\Gamma (\etapigg) = (0.58 \pm 0.30) $ eV.
This implies that the usual resonance saturation assumption may not be valid
for $O(p^{6})$ chiral lagrangian, but not as much as our result (12).
Again, only a new data can tell which approach is more suitable.

%\newpage

\vspace{.5in}

\newpage

{\Large \bf  Figure Captions}

\vspace{.3in}
\noindent
{\bf Figure 1}  The $m_{\gamma\gamma}^2$ spectrum in $\etapigg$ for
two values of $d_{3}$ : the solid curve for $d_{3} = - 1.4 \times 10^{-2}
{}~~{\rm GeV}^{-2}$ and the dashed curve for $d_{3} = 4.5 \times 10^{-2}
{}~~{\rm GeV}^{-2}$.  The dotted curve is for the vector meson exchanges only
({\it i.e.}, $ d_{3} = 0$).

%\vspace{.4in}
%\noindent
%{\bf Figure 2}  The reaction cross section  for $\ggpipi$ as a function of
%$m_{\gamma\gamma}$ for
%two values of $d_{3}$ : the solid curve for $d_{3} = - 1.5 \times 10^{-2}
%~~{\rm GeV}^{-2}$ and the dashed curve for $d_{3} = 4.0 \times 10^{-2}
%~~{\rm GeV}^{-2}$.  The dotted curve is for the vector meson exchanges only
%({\it i.e.}, $ d_{3} = 0$).  The data is taken by Crystal Ball
%Collaboration.
%
%\vspace{1.in}

\end{document}